\documentstyle[preprint,tighten,aps]{revtex}    
\begin{document}           %
\draft
\preprint{\vbox{\noindent
Submitted to Phys. Lett. B\hfill LGNS-94-104\\
          \null\hfill  INFNFE-9-94\\
          \null\hfill  INFNCA-TH-94-15}}
\title{The MSW solution to the solar neutrino problem \\
       for non-standard solar models. \\
      }
\author{
         V.~Berezinsky$^{(1)}$,
         G.~Fiorentini$^{(2)}$,
         and M.~Lissia$^{(3)}$
       }
\address{
$^{(1)}$INFN, Laboratori Nazionali del Gran Sasso, 67010 Assergi (AQ),
        Italy \\
$^{(2)}$Dipartimento di Fisica dell'Universit\`a di Ferrara, I-44100 Ferrara,
\\
        and Istituto Nazionale di Fisica Nucleare, Sezione di Ferrara,
        I-44100 Ferrara \\
$^{(3)}$Istituto Nazionale di Fisica Nucleare, Sezione
        di Cagliari, I-09128 Cagliari, \\ and
        Dipartimento di Fisica dell'Universit\`a di Cagliari, I-09100
        Cagliari
        }
\date{August 1994}
\maketitle                 
\begin{abstract}
The difficulties for non-standard solar models (NSSM) in resolving
the solar neutrino problem are discussed stressing the
incompatibility of the gallium--Kamiokande data, and
of the gallium--chlorine data.
We conclude that NSSM's cannot explain simultaneously the results of any
two of the solar neutrino data (chlorine, Kamiokande and gallium).
We address further the question whether
the MSW solution exists for NSSM's (e.g. models with
$^8$B neutrino flux much lower than the standard one
and/or central temperature $T_c$ very different
from $T_c^{\text{SSM}}$).
We demonstrate that the MSW solution exists and is very
stable relative to changes of $S_{17}$ ($S$-factor for
$p$ + $^7$Be reaction) and $T_c$. In particular, $\Delta m^2$ is almost
constant,
while $\sin^2 2\theta$ depends on the exact values of $S_{17}$
(or $^8$B-neutrino flux) and $T_c$.
\end{abstract}
\pacs{96.60.Kx, 14.60Pq}
\narrowtext
\section{Introduction}
\label{intro}
The standard solar model (SSM) gives a very good description of the Sun.
An impressive confirmation of the SSM is given by helioseismological
observations, which, according to Ref.~\cite{Turck93a},
agree with the SSM predictions at the level of
$0.5\%$ for distances down to $0.2R_{\odot}$.
As a matter of fact there are at least 12 SSM's whose predictions are in
reasonable
agreement~\cite{Bahcall88a,Lebreton88,Sackman90,Proffitt91,Guzik91,%
Bahcall92a,Guenter92,Christi92,Ahens92,Bertomieu93,Turck93b,Cast93b}.
All these models include the same
physics, and the slight differences between their results are mostly
caused by differences in the input parameters.

However, all SSM's predict neutrino fluxes which are in disagreement with the
observations of all four neutrino
experiments ~\cite{Davis94a,Davis94b,Kamio,Gallex,SAGE}
(see Table~\ref{Exp}).
This deficit
of the detected solar neutrinos is referred to as the solar neutrino
problem (SNP).

The SSM has uncertainties that basically reflect the uncertainties
in the input parameters. It has been
shown~\cite{Cast93a,Cast94b,Hata94a,Hata94b}
that the most important
uncertainties in the neutrino fluxes can be described by three parameters,
the central temperature $T_c$, the $S$-factor $S_{17}$ for the
$^7\text{Be} + p \to {}^8\text{B} + \gamma$ cross section,
and the ratio $S^2_{33}/S_{34}$ of the $S$-factors for
the reactions $^3\text{He} + ^3\text{He} \to ^4\text{He} + 2p$ and
$^3\text{He} + ^4\text{He}  \to ^7\text{Be} + \gamma$.
The uncertainty in the central temperature actually sums up uncertainties
in the astrophysical factor $S_{11}$ for the
$p+p \to ^2\text{He} + e^{+} + \nu$  cross section, in the
solar opacity, which depends in particular on the metal abundances
$Z/X$ and on possible collective plasma effects~\cite{Tsytovich},
in age of the Sun, and in some other quantities.

One can distinguish between SSM's and NSSM's.
All SSM's consider the same physical processes, and use
similar input parameters. On the contrary, NSSM's consider
large changes of the input parameters, often outside their estimated
uncertainties~\cite{Cast94b},
and/or introduce new physical processes.
In practice we assume that SSM's are characterized by the
input parameters and their uncertainties as given in Ref.~\cite{Cast94b},
which basically follows Bahcall~\cite{Bahcall92a,Bahcall89}.

Recently two new developments have attracted a great deal of attention.
If confirmed, they can change dramatically the prediction
for the solar neutrino fluxes.

The first one is the measurement~\cite{Moto94}
of $S_{17}$ factor from
the cross section of dissociation for the $^8$B nuclei in the Coulomb field of
$^{208}$Pb: $^8\text{B} + {}^{208}\text{Pb} \to {}^{208}\text{Pb}
+ {}^7\text{Be} + p $. The preliminary result
gives $S_{17}= 16.7 \pm 3.2$~eV~barn, which should be compared with
the value $S_{17}= 24 \pm 2$~eV~barn used by Bahcall and
Pinsonneault~\cite{Bahcall92a}. This new
result would imply a proportional reduction of the predicted $^8$B
neutrino flux, which would come close to the flux measured by Kamiokande.

The second new development consists in the theoretical consideration of
collective plasma effects in the Sun. These effects have the
potentiality of lowering the solar opacity by as much as
18\%~\cite{Tsytovich}, with the consequent lowering of the theoretical
prediction of the Sun central temperature by $2\div 3$\%.

The previous considerations have motivated us to study NSSM's where both
the astrophysical factor $S_{17}$ and the Sun central temperature are left
almost as free parameters. As discussed in the next Section,
this freedom, and even more so the aforementioned effects,
are insufficient to resolve the SNP. It is possible to
explain the observed $^8$B neutrino flux, i.e. the Kamiokande data,
but it is not possible satisfy any two of the three experimental data
(chlorine, Kamiokande and gallium) simultaneously.
The basic reason for this failure is that the SNP affects now also the
$^7$Be neutrinos, and not only the rare $^8$B
neutrinos~\cite{Cast94b,Bere94a,Bahcall94b}:
the combination of the two gallium experiments with
Homestake or Kamiokande, or the combination of Homestake with Kamiokande,
implies
a too low $^7$Be-neutrino flux.
As concluded in Sect.~\ref{noastro},
the astrophysical solution is strongly disfavored.

The MSW
 mechanism~\cite{MSW78,MSW86} offers the attractive possibility of
selective suppression of the $^7$Be neutrinos, and
reconciling the SSM with experiments. After having examined the input
parameter uncertainties and the NSSM's as possible
solutions to the SNP, it is only fair to examine how they affect the
MSW solution. In particular, we ask whether the MSW solution exists for
values of $S_{17}$ and central temperature significantly different from
the standard ones (the range of $S_{17}$ considered includes the new
preliminary result: $S_{17}=11\div23$~eV~barn~\cite{Moto94}).

In Section~\ref{MSW}, we shall demonstrate the validity of the MSW solution
and its weak dependence on the input parameters.
We find that considerable changes of the parameters $S_{17}$ and
$T_c$ affect the mixing angle,
but $\Delta m^2$ remains practically the same.

\section{Nuclear/Astrophysical solutions are disfavored}
\label{noastro}
At present the result of any (non-standard) solar model calculation, also
including uncertainties in the nuclear cross sections, is inconsistent not
only with the combination of the
chlorine and Kamiokande
data~\cite{Bahcall90a,Bahcall93a,Cast93a,Hata94a,Bere93a},
but also with the combinations of
the gallium data (GALLEX~\cite{Gallex} and
SAGE~\cite{SAGE})
with either the Kamiokande~\cite{Kamio} or chlorine~\cite{Davis94a,Davis94b}
data.
For a summary of the experimental data see
Table~\ref{Exp}.

The physical problem is the $^7$Be/$^8$B neutrino
ratio.
If both the chlorine and Kamiokande data
are correct within one standard deviation, the $^7$Be flux is negative. If we
allow both experiments to go two standard deviations away from their central
value, the $^7$Be flux must still be smaller than
$
1.85\times 10^9 \text{ cm}^{-2} \text{ sec}^{-1}
$,
i.e. less than 40\% of its SSM value.

A {\em model-independent} analysis along the lines of
Ref.~\cite{Cast94a}, which
only uses the luminosity constraint, shows that the combination of the
present data from the gallium experiments and from Kamiokande also
requires that the $^7$Be flux be smaller than
$
1.9\times 10^9 \text{ cm}^{-2} \text{ sec}^{-1}
$;
larger values can be obtained only going more than
two standard deviations from both data. These conclusions can be also
derived from the comparison
of the gallium and chlorine data.
A similar result was recently obtained by Bahcall~\cite{Bahcall94b}.

Nowadays the SNP is the problem of beryllium
neutrinos.
The low $^7$Be flux, implied by the experimental data, is eventually
the reason astrophysical solutions fail and, as we shall see,
the MSW solution is stable in mass.

At the $2\sigma$ level,
any two of the chlorine, Kamiokande and gallium data implies for all
existing SSM's that:
\begin{equation}
\label{Belim}
 \frac{\Phi(^7\text{Be})}{\Phi^{\text{SSM}}(^7\text{Be})} < 0.4 \quad .
\end{equation}
The diminishing of the astrophysical $S_{17}$ factor to the value
15--17~keV~barn can solve the boron neutrino problem,
but the flux of $^7$Be neutrino remains too
large for both the chlorine and gallium experiments.

In principle, we cannot exclude a strong suppression of
$^7$Be/$^8$B ratio
by nuclear physics effects, but we are able to obtain this effect only
at the price of large and correlated change of parameters, which eventually
are unrealistic. Let us give two examples.

Increasing the $S_{33}$ factor we can make both fluxes,
$^7$Be and $^8$B, {\em very} low.
However, in Ref.~\cite{Cast94a} it was demonstrated that increasing $S_{33}$
by factor 11.0 still gives a poor fit to the observational data.
To overcome this difficulty we can increase the proton-capture rate on
beryllium
$\lambda_p$, and/or decrease the electron capture rate $\lambda_e$.
In this manner we can increase the production of $^8$B neutrinos
proportionally to
$\frac{\lambda_p}{(\lambda_p+\lambda_e)}$ (therefore compensating for the
too large reduction caused by the increase of $S_{33}$), and
decrease the $^7$Be/$^8$B ratio as $\lambda_e/\lambda_p$.
However, this game works only with a
rather unrealistic combination of large correlated changes of two
($S_{33}$ and $\lambda_e$) or three ($S_{33}$, $\lambda_e$
and $\lambda_p$) parameters.

In Ref.~\cite{Cast94b} solar models with very generous changes of the input
parameters are studied. These input parameters include the central
temperature $T_c$ (which -- we recall -- is determined by heavy
element abundances,
$S_{11}$, solar age, new plasma effects, etc.), $S_{17}$ and $S^2_{34}/S_{33}$.
There it is demonstrated that arbitrary variations of any single one of
these parameters are insufficient to explain the data of all four experiments.

We can add to this demonstration the following proof of the impossibility of
an astrophysical solution where we vary
the central temperature, with the aim of suppressing the $^7$Be neutrino flux,
and, at the same time,
let the astrophysical factor $S_{17}$ be a free parameter, with the
hope of fixing the $^8$B neutrino flux.

The basic point is that the scale laws for the fluxes are such that
central temperature and/or $S_{17}$ variations yield:
\begin{eqnarray}
\label{Belaw}
\frac{\Phi(^7\text{Be})}{\Phi^{\text{SSM}}(^7\text{Be})}
&\approx&
\left( \frac{T_c}{T_c^{\text{SSM}}} \right)^{9\div 11} \\
\label{Blaw}
\frac{\Phi(^8\text{B})}{\Phi^{\text{SSM}}(^8\text{B})}
&\approx&
\frac{S_{17}}{S_{17}^{\text{SSM}}}
\left[
      \frac{\Phi(^7\text{Be})}{\Phi^{\text{SSM}}(^7\text{Be})}
\right]^2 \quad .
\end{eqnarray}
In fact the $^8$B production is linearly proportional to $S_{17}$, since
only a tiny fraction of the $^7$Be nuclei terminate through the
$^8$B branch, and, therefore,
the concentration of $^7$Be nuclei is practically
independent of $S_{17}$. The power of two in the $^7$Be flux comes about
because, while the $^7$Be flux varies with temperature roughly as
$T_c^{9\div 11}$, the $^8$B flux goes roughly as
$T_c^{18\div 22}$~\cite{Cast94b,Bahcall89}.
Therefore, Eqs.~(\ref{Belim}) and (\ref{Belaw})
imply that we need a
central temperature reduction of about 10\%, and
Eq.~(\ref{Blaw}) together with the fact that Kamiokande
sees at least 40\% of the SSM $^8$B flux implies that we need to increase
$S_{17}$ of at least  a factor 2.5.

A more careful analysis that takes into account all three
experiments
confirms this  result.
We made a correlated $\chi^2$ analysis as function
of the central temperature, and of the $S_{17}$ cross section. Since this
analysis is an extension of the one done in
Ref.~\cite{Cast94b}
for only two values
of $S_{17}$, we refer to it for
details on the correlation
matrix and the $\chi^2$ calculation.

Figure~\ref{fig1} shows
the iso-$\chi^2$ curves for (a) all data,
(b) only the gallium and Kamiokande data, and (c) only the gallium and
chlorine data. We do not present the analysis of the
Kamiokande and chlorine data,
which just confirms the well-know incompatibility.
We find the best $\chi^2$ for
the combination of gallium and chlorine data, as already seen in
Ref.~\cite{Cast94b}. But even if we consider this ``best'' combination,
we obtain a $\chi^2$ that has a 2\% probability of being a
statistical fluctuations only for
central temperatures at least 10\% lower than the
one predicted by the SSM, and $S_{17}$ larger than 60~eV~barn.
For all other combinations of data
we have not even been able of finding a $\chi^2$ compatible with
statistical fluctuations at the level of 2\%
in the range of temperatures shown.

The conclusion of
these calculations is that we are not able to construct
solar models that reproduce the experimental neutrino fluxes, even if
we arbitrarily disregard one of the experiment.
As a matter of fact most of the area in Fig.~\ref{fig1} does not
correspond to NSSM's, as the ones in Ref.~\cite{Cast94b}, but to
truly unrealistic models. In the example where we dial both
central temperature and $S_{17}$, for instance,
the cross section $S_{17}$ needs to be much too large compared with its
experimental
determinations~\cite{Kava69,Filippo83,Langa94,Moto94}.
Moreover, we obtain central
temperatures for which is not even possible to construct a solar
model~\cite{Cast94b}
(in these cases we have used power-law extrapolations of solar model
 calculation).

We see, therefore, that the combined results of all four neutrino experiments
are incompatible with the astrophysical solution. If one insists upon
this kind of solution, one must conclude that some of the
experiments are wrong. Following J.~Bahcall~\cite{Bahcall94b} one
can ask: ``How many solar neutrino experiments are wrong?''

One can resolve the Homestake/Kamiokande conflict by assuming that either
one is wrong. But then we are left with the conflict
between the remaining experiment and the gallium experiments. Should we
say that both the Kamiokande and chlorine experiments are wrong? or
is the gallium data together with either Homestake or Kamiokande that is
wrong?

We think that it is more reasonable to consider the non-standard neutrino
option, which can explain the results of all four neutrino experiments
simultaneously.
\section{The MSW solution}
\label{MSW}
In this Section we shall study the MSW solution for NSSM's.
We shall start with a semiquantitative explanation of the reason $\Delta m^2$
is practically the same for the large class of SSM's and NSSM's.

The MSW mechanism~\cite{MSW78,MSW86}
suppresses the $^7$Be neutrinos as long as they satisfy the resonant
condition in the Sun:
\begin{equation}
\label{highlim}
\Delta m^2 <
\frac{2\sqrt{2} G_F \rho E_{\text{Be}}}{\cos 2\theta } \approx
 \frac{12 (\text{meV})^2}{\cos 2\theta}  \quad .
\end{equation}
In this equation $E_{\text{Be}}=0.862$~MeV is the energy of the
$^7$Be neutrinos, and $\rho$ is approximately equal to the electron
density at the peak of the production region for the $^7$Be neutrinos
(we use 89\% of the central electron density).

Let us derive first the minimum $pp$ neutrino counting rate in the gallium
experiments. To this end we take the $2\sigma$ lower limit for gallium
experiments, $56.4$~SNU, and subtract from it the $^8$B and $^7$Be signals.
The $^8$B-neutrino contribution can be found from the
$2 \sigma$ upper limit of the Kamiokande counting rate. As already
said in the previous Section, the combination of any two neutrino data,
gives a $2\sigma$ upper bound on the $^7$Be neutrinos of
$
1.9\times 10^9 \text{ cm}^{-2} \text{ sec}^{-1}
$.
As a result we obtain  33~SNU as
a very stringent lower limit ($2\sigma$ away from all three neutrino data)
for the
$pp$ signal in the gallium experiments.
Therefore, we conclude that at least
half of the $pp$ signal survives. This condition results in the lower
limit for $\Delta m^2$:
\begin{equation}
\label{lowlim}
\Delta m^2 >
\frac{2\sqrt{2} G_F \rho E_{pp}}{\cos 2\theta } \approx
 \frac{4 (\text{meV})^2 }{\cos 2\theta} \quad .
\end{equation}
In this equation we use $\rho$  equal to the
67\% of the central electron density, and
$E_{pp}$ is defined in such way that contribution of neutrinos  with
energies less than $E_{pp}$ to a gallium detector is half of the total
$pp$ signal. Since most of the $pp$ signal in a gallium detector is produced by
neutrinos
near the
upper end of the energy spectrum (0.420~MeV) we use,
for the sake of discussion,  $E_{pp}= 0.410$~MeV.
Thus, if the MSW mechanism is the solution to the SNP,
$\Delta m^2$ is bounded in a narrow interval, given by
Eqs.~(\ref{highlim}) and (\ref{lowlim}), independently of the solar model
we use.

When $\sin^2 2\theta << 1$, the suppression is almost complete, and the
exact value of $\sin^2 2\theta$ determines only the width of
the energy window where neutrinos are suppressed. For very small mixing
angles, only the  beryllium line is suppressed; increasing
$\sin^2 2\theta$ we suppress more  $^8$B neutrinos and/or $pp$ neutrinos
(depending on the value of $\Delta m^2$).
Therefore variations of astrophysical $^8$B flux produce only variations of
the mixing angle, but leave $\Delta m^2$ unaffected.

Within the MSW solution the Kamiokande measurement
$\Phi(\text{B}) \geq 0.4 \, \Phi_{\text{SSM}}(\text{B})$,
gives the  minimum value which a solar model can predict.
For this flux we  need no suppression of the $^8$B neutrinos.
The energy window must be very narrow, and centered on the beryllium line.
The $\sin^2 2\theta$ has the smallest value.
For the opposite extreme case, the maximum value of the $\Phi(\text{B})$
reasonably allowed
by  the  parameter uncertainties  is about
$\Phi(\text{B}) \leq 1.5\, \Phi_{\text{SSM}}(\text{B})$.
This value implies a strong MSW suppression for
both Kamiokande and Homestake, which in this case observe about one third of
the predicted flux. Then,
a large angle solution, which give almost equal suppression for energy
above the resonance, becomes more favorable.

These qualitative results have been confirmed by a detailed
$\chi^2$ analysis of the MSW solutions as function of the astrophysically
produced
$^8$B flux, and of the central temperature.
The details of the calculations have already been reported in
Ref.~\cite{Cast94b}
for the case of the SSM $^8$B flux. Here we use the same
procedure for the two extreme cases of  minimal $^8$B flux compatible with
Kamiokande, and the maximal $^8$B flux allowed by the present
uncertainties, namely  2.09 and
$8.7 \times 10^6 \text{ cm}^{-2} \text{ sec}^{-1}$, respectively.
The results are shown in Fig.~\ref{fig2}a.

In agreement with the semi-quantitative analysis given above,
the reduced $^8$B flux allows
only the small angle solution (labeled by 0.4),
which is shifted towards smaller angles in comparison with the SSM solution
(labeled by 1). In this case, we obtain a relatively large
$\chi^2$ (5.0) for the
best fit. The reason is that this reduced flux is barely compatible with the
Kamiokande result. If we use a slightly larger flux, e.g. 0.5 of the SSM
value, we obtain  a $\chi^2$ similar to the other cases.

 When the $^8$B flux increases above the SSM value,
the large angle solution becomes more favorable. The allowed region becomes
larger and its  statistical  weight increases. Simultaneously
the small angle solution moves towards larger angles. For
an increment described by a factor 2 (not shown in the figure),
the best fit corresponds to the large angle solution.

Let us now examine the influence of the
other uncertainties
on the MSW solution. We parameterized these uncertainties by central
temperature, which affects both $^8$B and $^7$Be neutrino fluxes.
One might worry that this parameterization does not cover all
solar models, and does not describe the possible  correlations between
neutrino
fluxes~\cite{Gates94}.

A priori this criticism is correct. However, several numerical studies
have shown that, in practice, the effects of independent variations of
the metal fraction Z/X, the opacities,
the astrophysical factor $S_{pp}$, and the Sun age are well
reproduced by variation of the central
temperature of the Sun~\cite{Cast93a,Cast94b,Hata94a,Hata94b,Cast94a}.

Since the $^8$B flux depends much more strongly on the central temperature than
the $^7$Be flux, we expect that  effect of the temperature variation
 should be qualitatively similar to the variation of
the $S_{17}$ factor. We have performed the calculation varying $T_c$ within
the range
 $\pm 1\%$, which describes the uncertainties of the SSM and within the
range
 $\pm 2\%$,  which corresponds to NSSM's.
Results are shown in Fig.~\ref{fig3}. The MSW solution exists for all
values of temperatures considered. One can see the stability of
MSW especially as the mass range is concerned. The mixing angle
changes slightly
with temperature: the
cooler Sun prefers the smaller mixing angles, while the hotter Sun revives the
otherwise dying large angle solution.

We conclude this Section with an  issue which has been
recently raised in a  paper by Gates, Krauss and White~\cite{Gates94}. We
shall show that this effect does not change our conclusions.

The definition of the confidence level (C.L.) regions is always a problem
due both to our ignorance of the probability distribution,
and to the
different assumptions that one wants to test.
Our pragmatic approach is that we feel confident about conclusions
that do not depend too much on the C.L. definitions, while we
feel uneasy in the opposite case.
For this reason, we use in our calculations two different definitions for
95\% C.L. regions.

The first definition, marked (a) in Fig.~\ref{fig2}, answers the
following question: if the MSW solution with the given parameters exists,
what region of parameters has the
95\% probability of containing these true parameters? The maximum likelihood
procedure (with some extra assumptions) tells us that this region is given
by $\chi^2< \chi^2_{\text{min}} + \Delta\chi^2$, where $\chi^2_{\text{min}}$
is the value of $\chi^2$ for the best fit, and $\Delta\chi^2$
depends on the number of parameters and on the C.L. For example,
$\Delta\chi^2=5.99$ for 2 parameters and a 95\% C.L.
We used this definition in our previous publications~\cite{Cast94a,Fiore93}.

The second definition (b) answers a different question:
what is the region of parameters where the MSW mechanism does not contradict
the {\em experimental data} at a given C.L.?
Again under some additional assumptions, this region is given by
$\chi^2<  \delta\chi^2$, where now $\delta\chi^2$ depends on the number of
data points and the C.L. For example, $\delta\chi^2=7.82$
for 3 experiments and a 95\% C.L.

Note that definition (a) will always give us some confidence
region, since we assume that the MSW mechanism exists, while definition (b)
can yield no allowed region at all: in this case we  conclude that the
MSW mechanism fails to provide the given C.L. In fact, only when
$\chi^2_{\text{min}}$ is large (bad fit) the two definitions give qualitatively
different regions: in this case we should worry about the original assumption
under which (a) is valid, i.e. that the MSW solution exists.

Figure~\ref{fig2} shows that for our purposes the two definitions are
very similar, and cannot change our conclusions.
The same is true for all our calculations, and, therefore,
in Fig.~\ref{fig3} we have used only the first definition for the confidence
regions.
\section{Conclusions}
\label{conclu}
For standard massless neutrinos the data of any two solar neutrino
measurements (chlorine, Kamiokande and gallium) are incompatible. The
incompatibility of the chlorine and Kamiokande data is a well recognized
problem~\cite{Bahcall90a,Bahcall93a,Cast93a,Hata94a,Bere94a,Bere93a}.
The conflict between the gallium and Kamiokande data, and between the
gallium and chlorine data,
can be shown in different ways. One can
obtain~\cite{Bere94a}
a rigorous
model-independent lower limit for the gallium detector counting rate by
neglecting the $^7$Be neutrino flux, using the observational lower limit
for the $^8$B neutrino flux, and taking the flux of $pp$ neutrinos from the
solar luminosity restriction. It gives $82.5$~SNU to be compared with the
combined result of the two gallium experiments $74 \pm 9$~SNU.
Using the gallium data, the luminosity constraint, and
the lower limit for the flux of $^8$B
neutrinos from either Homestake or Kamiokande,
one can derive~\cite{Cast94b}
a model-independent limit on the $^7$Be neutrino flux. The updated limit using
the new gallium
data~\cite{Gallex,SAGE}
is
$1.9 \times 10^9~\text{cm}^{-2}\text{s}^{-1}$ ($2\sigma$),
i.e. less than $40\%$ of the SSM value.
Most recently J.~N.~Bahcall~\cite{Bahcall94b}
found that the $^7$Be neutrino signal in a gallium detector is less
than 19~SNU at the 95\% C.L., about half of the
36~SNU predicted by the SSM.

We have demonstrated that there are no values of $S_{17}$ and $T_c$,
the two most important solar-model parameters as neutrino fluxes are
concerned, which satisfy
the combination of any two experimental data out of three
(chlorine, Kamiokande and gallium).

Nowadays, the essence of solar neutrino problem is the low $^7$Be neutrino
flux. Nuclear/astrophysical solutions to the solar neutrino problem are
strongly disfavored. The MSW mechanism offers a very attractive solution
to the solar neutrino problem.

The new measurements of $S_{17}$ factor~\cite{Moto94}
and collective plasma
effects~\cite{Tsytovich}
can considerably change the solar-model parameters $S_{17}$
and $T_c$. In fact, we might face what now are non-standard solar
models, which are also incompatible with the solar neutrino data.
Does the MSW solution exist for these models?

We have demonstrated that the MSW solution exists for NSSM's with
parameters $S_{17}$ and $T_c$ within their realistic uncertainties, and beyond.
Our conclusion concerning the $^8$B neutrino flux
coincides with the results of Ref.~\cite{Krastev94}.
The MSW solution is stable in mass ($\Delta m^2$)
even if $S_{17}$ or $T_c$ are drastically changed (a change of $S_{17}$
practically results in only a proportional change of the $^8$B  neutrino
flux). In particular,
$\Delta m^2$ is restricted to the range
$ 4 (\text{meV})^2<\Delta m^2 < 12 (\text{meV})^2$.
The physical reason of this
stability is the fact that the solar neutrino experiments observe most of
the predicted flux of $pp$ neutrinos,
and very small fraction of the $^7$Be neutrino flux.

Solar-model predictions for the $^8$B neutrino flux are only important
for the determination of the mixing angle.
Models with reduced $^8$B neutrino flux prefer the small
angle solution, because of the stronger suppression of the
$^7$Be/$^8$B flux ratio. The large angle solution disappears for models with
reduced $^8$B neutrino flux. Large-mixing-angle solutions reappear
in models with hotter core and/or larger
$S_{17}$ factor, since they predict a $^8$B-neutrino-flux
increase
larger than the corresponding increase
of the $^7$Be flux.

\acknowledgments
One of us (ML) wishes to thank the Organizers of the  ``Summer Institute
on Nuclear Physics and Astrophysics: Prospects for Underground Research''
held at Laboratori Nazionali del Gran Sasso, Italy, in June and July 1994,
where part of this work was done.
\section*{Note added}

While this paper was in preparation we received two
preprints~\cite{Bahcall94b,Krastev94}
that are relevant to our work.

J.~N.~Bahcall~\cite{Bahcall94b}
studies the incompatibility of the gallium data with either
Homestake or Kamiokande. Our method is essentially
different from his, but our conclusions coincide.
This same incompatibility was also discussed in
Refs.~\cite{Cast94b}~and~\cite{Bere94a}.

P.~I.~Krastev and
A.~Yu.~Smirnov~\cite{Krastev94}
obtain some of our same results. In particular,
these authors notice the mass stability of the MSW solution
relative to variation of $^8$B neutrino flux. Numerically, our results
agree with theirs.

\begin{figure}
\caption[chi2_ts]{
   Contours of equal $\chi^2$ for the neutrino fluxes in solar models
   parameterized
   by the central temperature (normalized to its SSM value), and by
   the $S_{17}$ astrophysical factor.
   Full contours correspond
   to $\chi^2$ equal to 40, 30, 20 and 10; dashed contours correspond
   to $\chi^2$ equal to 35, 25, 15 and 5. Values of $\chi^2 > 9.84$
   have a 2\%
   probability for three data (a), i.e. for all experiments
   (the two gallium data have been combined).
   Cases (b) and (c) show $\chi^2$ for two experimental data:
   gallium and Kamiokande
   (b) and gallium and chlorine (c). In this case values of
   $\chi^2 > 7.82$ have a 2\% probability. Graphs
   (b) and (c) show the contradiction between the gallium results,
   and either of two other experiments (chlorine or Kamiokande):
   even a $\chi^2 < 10$ (less than 1\% probability for values
   larger than 10) needs too high values of $S_{17}$.
               }
\label{fig1}
\end{figure}
\begin{figure}
\caption[boro]{
 Regions of the MSW solution allowed at the 95\% C.L. The region labeled 1
 corresponds to the SSM~\protect\cite{Cast93b} with
 the standard $^8$B neutrino flux; in the region labeled 0.4 (1.5)
 the $^8$B flux is reduced (increased) by a factor 0.4 (1.5).
 Note that the large angle solution disappears when the flux is
 reduced (0.4).
 The filled black circles show the best fit for each flux: the
 corresponding $\chi^2_{\text{min}}$ are 5.0, 0.22 and 0.32, for
 flux factors 0.4, 1 and 1.5, respectively.
 The confidence regions are defined as
 (a) $\chi^2= \chi^2_{\text{min}} + 5.99$, and
 (b) $\chi^2= 7.82$ (see text).
                }
\label{fig2}
\end{figure}
\begin{figure}
\caption[temp]{
 Regions of the MSW solutions allowed at the 95\% C.L. for models with
 varied central temperature. In (a) the central temperature $T_c$ is
 $\pm 1\%$ of $T_c^{\text{SSM}}$, the corresponding temperature in the
 SSM~\protect\cite{Cast93b}.
 Note that the large angle MSW solution disappears when the temperature is
 reduced. In (b) the central temperature
 is $\pm 2\%$ of $T_c^{\text{SSM}}$ (these are already NSSM's). There is
 again no large
 angle solution when the temperature is reduced. In both
 figures the black filled circles give the best fit:
 the corresponding $\chi^2_{\text{min}}$ are 1.38, 0.76, 0.22, 0.12 and 0.45,
 respectively for temperature changes of
 -2\%, -1\%, no change (SSM), +1\% and +2\%.
 The confidence regions are defined as
 $\chi^2= \chi^2_{\text{min}} + 5.99$ (see text).
                }
\label{fig3}
\end{figure}
\begin{table}
\caption[xman]{
         Most recent
         experimental data (Experiment), and a selected sample of
         theoretical predictions:
         Bahcall and Pinsonneault~\protect\cite{Bahcall92a} (BP SM),
         Turck-Chi\`eze {\it et al.}~\protect\cite{Turck93b} (TCL SM)
         and
         Castellani {\it et al.}~\protect\cite{Cast94a} (CDF SM). In this
         work we combine systematic and statistical errors
         of the experimental data and the GALLEX and SAGE data,
         and use a slightly different chlorine data (Data used).
         The errors of the theoretical works are
         1$\sigma$ effective errors.
         \label{Exp}
         }
\begin{tabular}{lr@{}l@{}lr@{}l@{}l@{}lr@{}l@{}lr@{}l@{}l}
&\multicolumn{3}{c}{Chlorine}&
  \multicolumn{4}{c}{Kamiokande}&
  \multicolumn{6}{c}{Gallium (SNU)}\\
&\multicolumn{3}{c}{(SNU)}&
  \multicolumn{4}{c}{(10$^6$ cm$^{-2}$ s$^{-1}$})&
  \multicolumn{3}{c}{GALLEX}&
  \multicolumn{3}{c}{SAGE}\\
\tableline
Experiment                    &2.&32 &$\pm$0.23%
\tablenote{Ref.~\protect\cite{Davis94b}}
                                          & 2.&90 &$\pm$0.23&$\pm$0.35%
\tablenote{Ref.~\protect\cite{Kamio}}
                                                &  79 &$\pm$10&$\pm$6%
\tablenote{Ref.~\protect\cite{Gallex}}
                                                &  69 &$\pm$11&$\pm$6%
\tablenote{Ref.~\protect\cite{SAGE}}
                                                                             \\
Data used                        &2.&28 &$\pm$0.23%
\tablenote{Ref.~\protect\cite{Davis94a}}
                                          & 2.&90 &$\pm$0.42&%
                               &\multicolumn{6}{c}{$74.4\pm 8.8$} \\
BP SM
                         &8.&0  &$\pm$1.0 & 5.&69 &$\pm$0.80&
                               &\multicolumn{6}{c}{132$^{+7}_{-6}$} \\
TCL SM
                         &6.&4  &$\pm$1.4 & 4.&4  &$\pm$1.1&
                               &\multicolumn{6}{c}{123$\pm$ 7} \\
CDF SM
                         &7.&8  &$\pm$1.0 & 5.&6  &$\pm$0.8&
                               &\multicolumn{6}{c}{130$\pm$ 6} \\
\end{tabular}
\end{table}

\begin{thebibliography}{10}
\bibitem{Turck93a}
S.~Turck-Chi\`eze et al., Phys. Rep. 230 (1993) 57.

\bibitem{Bahcall88a}
J.~N. Bahcall and R.~K. Ulrich, Rev. Mod. Phys. 60 (1988) 297.

\bibitem{Lebreton88}
Y.~Lebreton and W.~D\"{a}ppen, in: Seismology of the Sun and the
   Sun-like Stars, eds. V.~Domingo and E.~J.~Rolfe (European
   Space Agency, Nordwijk, 1988).

\bibitem{Sackman90}
I.~J.~Sackman, A.~I.~Boothroyd and W.~A.~Fowler, Ap.~J. 360 (1990) 727.

\bibitem{Proffitt91}
C.~R.~Proffitt and A.~N.~Cox, Ap.~J. 380 (1991) 238.

\bibitem{Guzik91}
J.~A.~Guzik and A.~N.~Cox, Ap.~J. 381 (1991) 331.

\bibitem{Bahcall92a}
J.~N. Bahcall and M.~H. Pinsonneault, Rev. Mod. Phys. 64 (1992) 885.

\bibitem{Guenter92}
D.~B.~G\"{u}nter et al., Ap.~J. 387 (1992) 372.

\bibitem{Christi92}
J.~Chistiansen-Dalsgaard, Geophys. Astrophys. Fluid. Dyn. 62 (1992) 123.

\bibitem{Ahens92}
B.~Ahrens, M.~Stix and M.~Thorn, A.\& A. 264 (1992) 673.

\bibitem{Bertomieu93}
G.~Bartomieu, J.~Provost and J.~Morel, A.\& A. 268 (1993) 775.

\bibitem{Turck93b}
S.~Turck-Chi{\`{e}}ze and I.~Lopes, Astrophys. J. 408 (1993) 347.

\bibitem{Cast93b}
V.~Castellani, S.~Degl'Innocenti and G.~Fiorentini,  A.\& A. 271 (1993) 601.

\bibitem{Davis94a}
R.~Davis Jr., in: Proc. of the 23rd ICRC (Calgary, Canada 1993), Prog.
  in Nucl. and Part. Phys. 32 (1994).

\bibitem{Davis94b}
R.~Davis Jr., Report on the Homestake solar neutrino experiment, in:
  Proc. of the International Topical Workshop on ``Solar Neutrino Problem:
  Astrophysics or Oscillations?'' (Assergi, Italy, February 1994),
  Vol. 1, eds.  V.~Berezinsky and E.~Fiorini (LNGS, 1994) p. 66.

\bibitem{Kamio}
A.~Suzuki, Kamiokande results and prospects, in: Proc. of the 6th
  International Symposium on Neutrino Telescopes (Venice, February 1993),
  ed. M. Baldo Ceolin.

\bibitem{Gallex}
GALLEX Collaboration, P.~Anselman et al., Phys. Lett. B 327 (1994) 377.

\bibitem{SAGE}
SAGE collaboration, J.~N.~Abdurashitiv et al., in: Proc. of the 27th
Int.~Conf. on High Energy Physics (Glasgow, July 1994).

\bibitem{Cast93a}
V.~Castellani, S.~Degl'Innocenti and G.~Fiorentini,  Phys. Lett. B 303
   (1993) 68.

\bibitem{Cast94b}
V.~Castellani et al., Ferrara Preprint INFNFE-3-94 (1994),
                      to appear in: Phys.~Rev.~D.

\bibitem{Hata94a}
N.~Hata, S.~A.~Bludman and P.~Langacker, Phys. Rev. D 49 (1994) 3622.

\bibitem{Hata94b}
N.~Hata, University of Pennsylvania Preprint UPR-0612T (1994).

\bibitem{Tsytovich}
V.~Tsytovich, Collective plasma effects in the radiative transport in
              solar interior, in: Proc. of the International
  Topical Workshop on ``Solar Neutrino Problem:
  Astrophysics or Oscillations?'' (Assergi, Italy, February 1994),
  Vol. 1, eds.  V.~Berezinsky and E.~Fiorini (LNGS, 1994) p. 238.

\bibitem{Bahcall89}
J.~N. Bahcall, Neutrino Astrophysics (Cambridge University Press,
  Cambridge, 1989).

\bibitem{Moto94}
T.~Motobayashi et al., Rikkyo Preprint RUP-94-2  (1994).

\bibitem{Bere94a}
V.~Berezinsky, INFN LNGS Preprint LNGS-94/101, to appear in:
               Proc. of the 7th International Symposium on Neutrino
               Telescopes (Venice, February 1994),
               ed. M. Baldo Ceolin.

\bibitem{Bahcall94b}
J.~N.~Bahcall, Princeton IAS Preprint IASSNS-AST-94-37 (1994).

\bibitem{MSW78}
L. Wolfenstein, Phys. Rev. D 17, (1978) 2369.

\bibitem{MSW86}
S.~P. Mikheyev and A.~Yu.~Smirnov, Nuovo Cimento C 9 (1986) 17.

\bibitem{Bahcall90a}
J.~N.~Bahcall and H.~A.~Bethe, Phys. Rev. Lett. 65 (1990) 2233.

\bibitem{Bahcall93a}
J.~N.~Bahcall and H.~A.~Bethe, Phys. Rev. D 47 (1993) 1298.

\bibitem{Bere93a}
V.~Berezinsky, INFN LNGS Preprint LNGS-93/86 (1993),
               to appear in: Comments on Nuclear and
               Particle Physics.

\bibitem{Cast94a}
V.~Castellani et al., Phys. Lett. B 324 (1994) 245.

\bibitem{Kava69}
R.~W.~Kavanagh et al., Bull. Am. Phys. Soc. 14 (1969) 1209.

\bibitem{Filippo83}
B.~W.~Filippone et al., Phys. Rev. C 28 (1983) 2222.

\bibitem{Langa94}
K.~Langanke and T.~D.~Shoppa, Cal Tech Kellogg Lab Preprint MAP-168 (1994).

\bibitem{Gates94}
E.~Gates, L.~M.~Krauss and M.~White, Fermilab Preprint FERMILAB-PUB-94-176-A
               (1994).

\bibitem{Fiore93}
G.~Fiorentini et al., Phys. Rev. D 49 (1994) 6298.

\bibitem{Krastev94}
P.~I.~Krastev and A.~Yu.~Smirnov, Preprint DOE-ER-40561-137  and INT94-13-02
               (1994).
\end{thebibliography}
\end{document}